\documentclass[%
 reprint,
 amsmath,amssymb,
 aps,
]{revtex4-2}
\usepackage{xcolor}
\usepackage{array}  %
\usepackage{graphicx}
\usepackage{dcolumn}
\usepackage{bm}
\usepackage{booktabs}  
\usepackage{siunitx} 
\usepackage{multirow}
\usepackage[version=4]{mhchem}

\definecolor{DarkRed}{rgb}{0.6,0,0}

\begin{document}

\title{Polaron-mediated anisotropic exchange in 2D magnets}

\author{Johanna P. Carbone$^{1}$}
\email{johanna.paulina.carbone@univie.ac.at}
\author{Jakob Baumsteiger$^{1,2,3}$}
\author{Cesare Franchini$^{1,3}$}

\affiliation{$^{1}$Faculty of Physics and Center for Computational Materials Science, University of Vienna, Kolingasse 14-16, 1090 Vienna, Austria \\
$^{2}$Vienna Doctoral School in Physics, University of Vienna,ex Boltzmanngasse 5, 1090 Vienna, Austria\\
$^{3}$Department of Physics and Astronomy “Augusto Righi”, Alma Mater Studiorum – Università di Bologna, 40127 Bologna, Italy}

\begin{abstract}
Two-dimensional (2D) magnets offer a rich platform for exploring emergent spin phenomena due to their unique and diverse magnetic properties. Beyond intrinsic magnetism, external manipulation—such as defect engineering, molecular adsorption, or charge doping—offers powerful routes to control their magnetic behavior. In this work, we demonstrate that localized electron polarons provide an effective means to modulate magnetism in 2D magnets. Using first-principles  calculations, we investigate polaron formation in monolayer MnPS$_3$
and compute the resulting changes in magnetic exchange interactions. Our results reveal that polarons can locally break magnetic symmetry and induce anisotropic exchange couplings, highlighting a novel mechanism for tuning magnetic textures. This insight opens promising pathways for designing atomic-scale control of magnetism, with potential impact on spintronic technologies.
\end{abstract}

\maketitle

\section{Introduction}

2D magnetic materials have attracted significant attention in recent years, as their wide range of behaviors opens up new possibilities for the development of efficient and controllable next-generation magnetic devices. Notable examples include transition-metal halides and chalcogenides such as CrI$_3$, VI$_3$, Fe$_3$GeTe$_2$, and NiPS$_3$, among many others \cite{gibertini2019magnetic,kumari2021recent,wang2022magnetic}.
This interest is largely driven by the variety of intrinsic properties that emerge in 2D systems, including a broad spectrum of electronic phases—from semiconducting to metallic—alongside spin-orbit coupling effects, topological states, and complex magnetic textures that can markedly differ from their bulk counterparts due to reduced dimensionality and enhanced quantum confinement \cite{novoselov20162d,ahn20202d,wei2020emerging,culcer2020transport,premasiri2019tuning,frey2018tuning}.

Further promoting this rapid development are recent advances in the fabrication of 2D magnets, such as mechanical exfoliation, molecular beam epitaxy, and chemical vapor deposition \cite{rahman2021recent,hossain2022synthesis}, which have made it possible to isolate atomically thin magnetic layers. Crucially, the demonstration of long-range magnetic order at finite temperatures—enabled by magnetocrystalline anisotropy—has challenged the limitations imposed by the Mermin–Wagner theorem \cite{mermin1966absence}, thereby allowing stable magnetism in 2D systems.

While intrinsic magnetism in 2D materials is already rich and complex, an additional route to manipulate and control their magnetic properties is through external means—such as the introduction of defects \cite{cheng2018robust,mayoh2021effects,zhao2018surface}, adatoms \cite{kim2019exploitable,song2019tunable,guo2018half}, or adsorbed molecules \cite{tang2020tunable,he2019remarkably,rassekh2020remarkably}.
In MnPS$_3$ monolayers, it has been shown that molecular adsorption breaks spatial inversion symmetry, leading to the emergence of a finite Dzyaloshinskii–Moriya interaction \cite{wang2022impacts}. In addition to symmetry breaking, adsorption induces significant structural and electronic modifications that alter the Mn–S–Mn superexchange interactions. These changes, in turn, affect both the magnetocrystalline anisotropy and the isotropic exchange interactions. 

An alternative approach to controlling magnetic properties is charge doping, which can induce or modify magnetic ordering and tune the strength of magnetic interactions, as recently demonstrated in CrI$_3$ \cite{Orozovic2025}. A particularly remarkable consequence of charge doping is the formation of polarons—composite quasiparticles arising from the interaction between excess charge carriers and the phonon field \cite{franchini2021polarons,emin2013polarons}, which have also been observed in atomically thin crystals~\cite{Sio2023,liu2023atomic,cai2023manipulating,Yao2024}. Since polarons are accompanied by anisotropic lattice distortions, it is plausible that their formation could significantly modify magnetic interactions through the associated structural and electronic rearrangements.


While polarons have been extensively studied in bulk polarizable crystals~\cite{franchini2021polarons} including magnetic systems \cite{Rho2002,celiberti2024spin,Redondo2024}, their role in 2D magnetic materials remains relatively unexplored. Only a few studies have begun to address this emerging field.  For instance, in monolayer CoCl$_2$, single electron polarons were created by injecting electrons using a scanning tunneling microscope (STM) through a voltage pulse \cite{liu2023atomic,cai2023manipulating}. It was shown that polarons in the system can be also actively created, erased, and repositioned by the electric field generated by the STM tip. 
However, the impact of these localized charges on the magnetic properties of the system was not addressed.

In this work, we aim to fill this gap by 
studying how localized electron polarons affect magnetic exchange interactions in MnPS$_3$, an experimentally exfoliable \cite{neal2019near} 2D antiferromagnet proposed as a candidate for spin current generation via the spin photogalvanic effect \cite{xiao2021spin}. Our findings reveal a clear connection between polaron formation and the emergence of anisotropic magnetic coupling, pointing toward new directions for functional control of magnetism at the atomic scale.
As a consequence, self-trapped charges in 2D magnets can locally alter the magnetic texture, potentially giving rise to non-collinear magnetic configurations with promising implications for spintronic and magnetoelectric applications \cite{liu2024spin,rimmler2025non,qin2020noncollinear,jungwirth2018multiple}. 

MnPS$_3$ features a monolayer structure composed of a hexagonal lattice of Mn$^{2+}$ cations, each octahedrally coordinated by sulfur atoms. Positioned at the center of each Mn hexagon is a $[\text{P}_2\text{S}_6]^{4-}$ bipyramidal unit, oriented perpendicular to the plane. This arrangement gives rise to a primitive unit cell containing two Mn atoms, two P atoms, and six S atoms. As a result of this structural configuration, MnPS$_3$ behaves as an antiferromagnetic (AFM) semiconductor, with a magnetic ground state characterized by Néel-type ordering \cite{gao2023photoexcitation,wang2022impacts,samal2021two,yang2020electronic}. \\

\section{Theoretical methods and computational setup}

We have addressed the problem by means of first principles calculations within the spin non-collinear DFT+$U$ framework using the Vienna ab initio Simulation Package (VASP)~\cite{kresse1996efficient}. { A $3\times3\times1$ supercell (total of 90 atoms) is employed to model the MnPS$_3$ monolayer both in its pristine form and in the presence of a localized polaron. The calculations employ the Perdew–Burke–Ernzerhof (PBE) exchange–correlation functional~\cite{perdew1996generalized}, with a plane-wave energy cutoff of 600 eV and a $3\times3\times1$ $k$-point mesh used for the self-consistent field calculations.}

{Within the DFT+$U$ approach with a Hubbard parameter of $U = 5$ eV applied to the Mn $3d$ states, 
the band gap of MnPS$_3$ has been reported as $E_g = 2.34$ eV \cite{gao2023photoexcitation}, $E_g = 2.40$ eV \cite{wang2022impacts}, and $E_g = 2.50$ eV in \cite{yang2020electronic}, with the latter also reporting a higher value of $3.25$ eV obtained from calculations using the HSE06 hybrid functional. This range of band gap values provides a sufficiently wide energy window to support the formation and stabilization of a localized polaronic state. Thus, to account for the strong on-site Coulomb interactions of the Mn $3d$ electrons, we use a Hubbard $U$ parameter of $5$ eV following the Dudarev approach \cite{dudarev1998electron}. With this set-up, we determined the lattice constant of the relaxed unit cell to be $6.14$ \AA \ and a magnetic moment of $m_s=4.6 \ \mu_{\rm{B}}$ on the Mn atoms.}

To extract the magnetic exchange interactions, the energy mapping four-states method is employed \cite{picozzi2024spin,li2021spin,xiang2011predicting}. In this method, each component $(a,b)$ of the exchange interaction tensor $J_{ij}$, associated with a specific atomic pair $(i,j)$, is calculated using the expression:
\begin{equation}
    J_{ij}^{ab} = \frac{E_{ij,\uparrow\uparrow}^{ab} + E_{ij,\downarrow\downarrow}^{ab} - E_{ij,\uparrow\downarrow}^{ab} - E_{ij,\downarrow\uparrow}^{ab}}{4S^2},
\end{equation}
where $(a,b)$ denote the Cartesian components ($x$, $y$, $z$) and $S$ is the magnitude of the spin moment on sites $i$ and $j$. 
The quantities $E_{ij}^{ab}$ represent the total energies computed for the four distinct spin configurations of the pair: parallel ($\uparrow\uparrow$, $\downarrow\downarrow$) and antiparallel ($\uparrow\downarrow$, $\downarrow\uparrow$). 

\begin{figure}[t!]
 \hspace{-1cm}
    \includegraphics[scale=0.58]{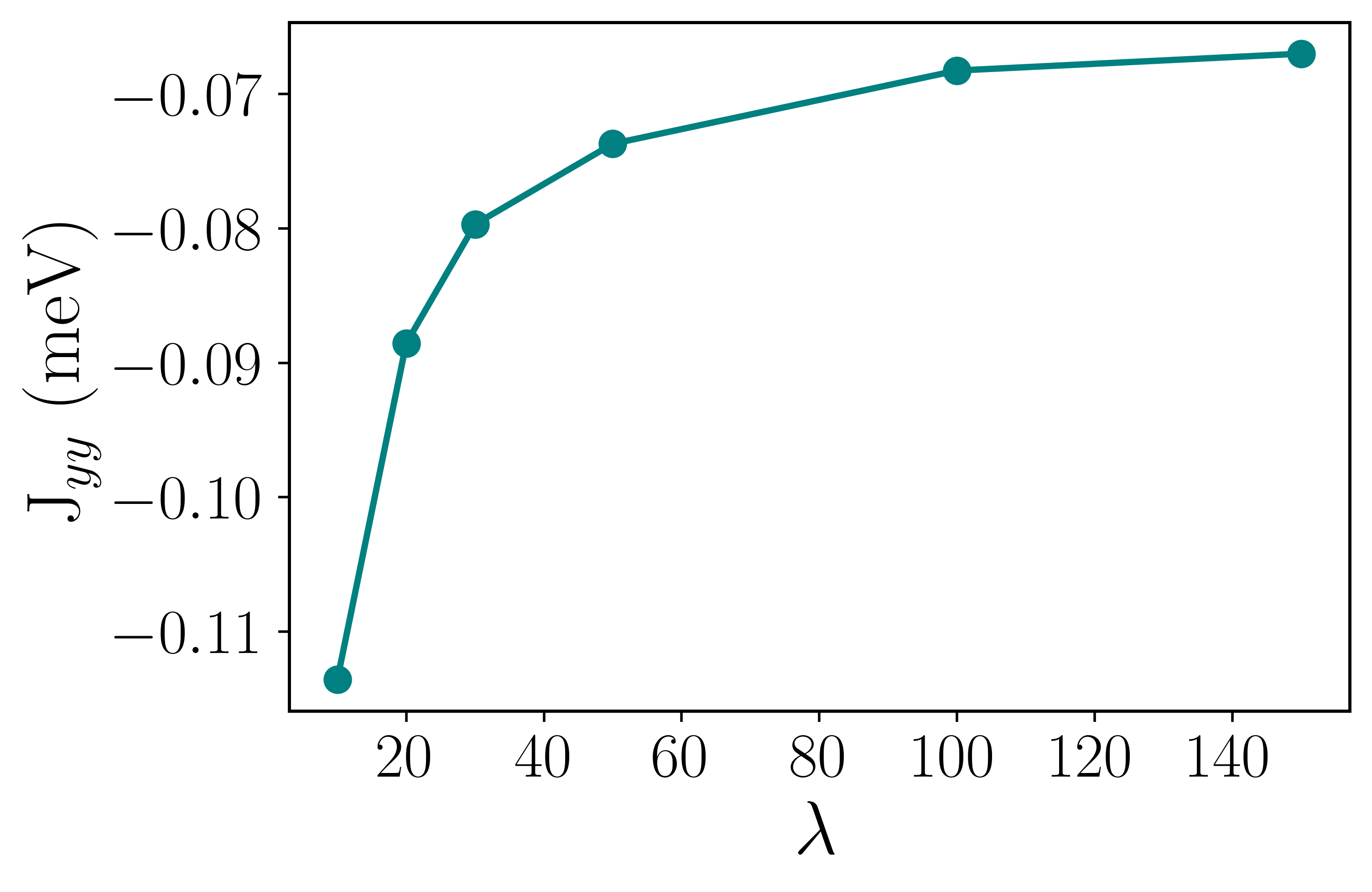}
    \caption{Convergence of the $J_{1,2}^{yy}$ component of the polaronic system with the penalty parameter $\lambda$.}
    \label{Figure4}
\end{figure}

{To determine the energies $E_{ij}^{ab}$, we constrained both the direction and sign of the magnetic moments. The total energy in this approach is given by
\begin{equation}
    E = E_0 + \sum_{I} \lambda \left( |\vec{M}_{I}| - \hat{M}_I^0 \cdot \vec{M}_{I} \right),
\end{equation}
where $E_0$ is the DFT energy, and the second term represents a penalty energy. Here, $\vec{M}_{I}$ is the magnetic moment obtained by integration over a sphere of specified volume at site $I$, and $\hat{M}_I^0$ is the target magnetic moment at the same site. The parameter $\lambda$ determines the weight of the penalty term in the total energy and can be adjusted as needed.}
{Since the magnitude of the smallest exchange component is on the order of $10^{-5}$~eV, we performed a convergence test of this component with respect to the penalty parameter (Figure~\ref{Figure4}). A value of $\lambda = 100$ was chosen to enforce the rotation of the magnetic moments, as the exchange tensor component $J_{12}^{yy}$ in the polaronic case was observed to converge around this value.
For the polaronic system, we also recalculated the $J_{2,3}$ tensor including spin-orbit coupling and found that it did not lead to any noticeable difference.}\\

\section{Results and discussion}

\subsection{Polaron formation}

\begin{figure*}[t!]
    \centering
    \includegraphics[width=\textwidth]{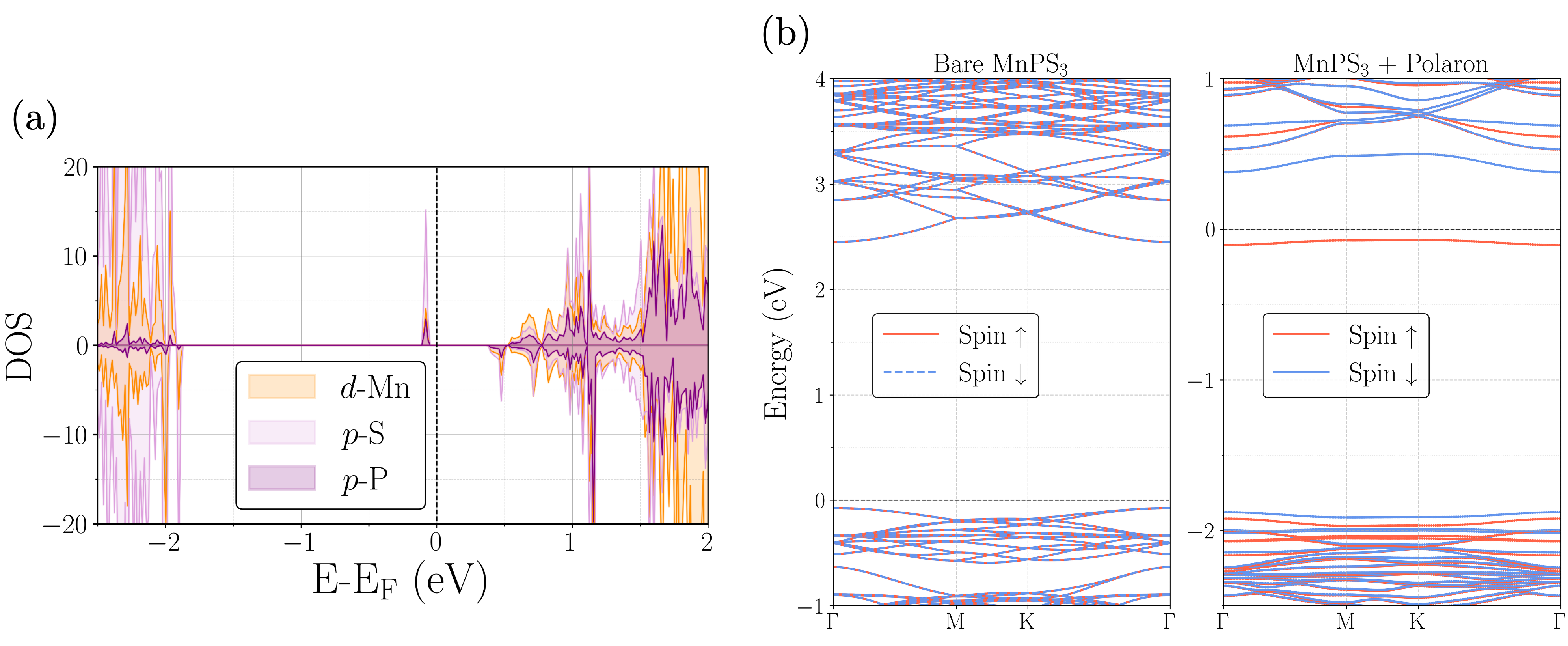}
    \caption{(a)  Spin-polarized projected DOS of MnPS$_3$ monolayer, showing Mn $d$ states (orange), P $p$ states (purple), and S $p$ states (pink). The upper panel (DOS $>$ 0) represents the spin-up channel, while the lower panel (DOS $<$ 0) corresponds to the spin-down channel. (b) Spin-polarized band structure of the bare, undoped $3\times3\times1$ MnPS$_3$ monolayer and of the polaronic solution. The spin-up and spin-down channels are shown in red and blue, respectively. Note that the two band structures span different energy ranges.}
    \label{Figure1}
\end{figure*}
{To assess the stability of a polaron within the system, we calculated its formation energy, which is defined as the energy difference between a localized polaronic state and a delocalized charge state. Specifically, it is calculated by subtracting the total energy of the polaronic configuration, where the excess electron occupies a localized in-gap state accompanied by a lattice distortion, from the total energy of the delocalized configuration, in which the excess charge is spread across the conduction band minimum without significant atomic rearrangement \cite{reticcioli2019small}. 
This energy difference captures both the electronic stabilization due to charge localization and the elastic energy cost associated with the accompanying lattice deformation.} 

To obtain the polaronic solution, we employ a two-step procedure. First, a spin-polarized ionic relaxation is performed by adding one extra electron to the system, introducing small initial distortions in the atomic positions, and applying a large Hubbard $U$ value to promote electron localization. In the second step, the Hubbard $U$ is removed and the structure is further relaxed. We modeled the polaron using the approach described above, initializing the excess charge on a P atom, and present the corresponding results below. Initialization on an S or Mn atom leads to the same polaronic configuration.
\begin{figure*}[t!]
    \centering
    \includegraphics[width=\textwidth]{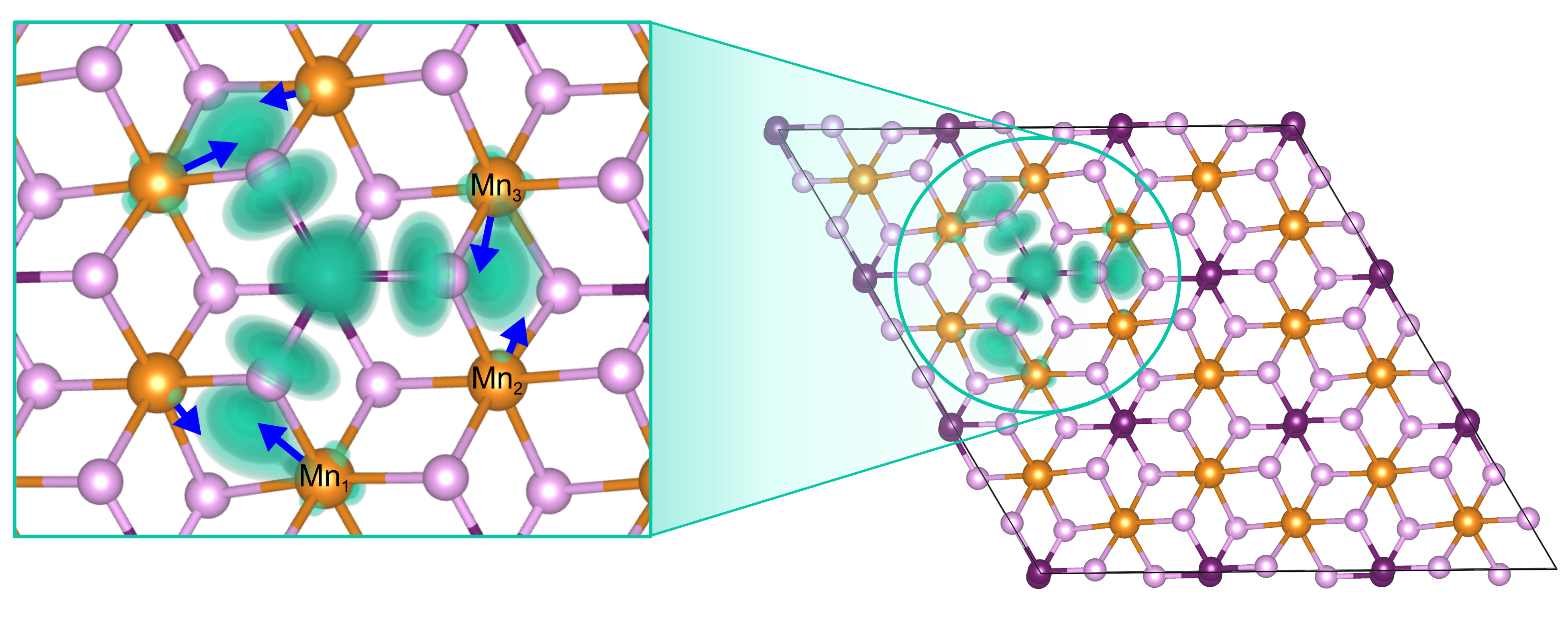}
    \caption{
    Crystal lattice of $3\times3\times1$ \ce{MnPS3} monolayer with a polaron, visualized as the torquoise charge-density cloud corresponding to the in-gap peak in Figure~\ref{Figure1}. The displacement of Mn atoms induced by the polaron is indicated by blue arrows in the close-up figure. The Mn atoms that were used for the calculation of the nearest-neighbor exchange matrices $J_{ij}$ are labelled. Visualization done using \textsc{vesta} \cite{momma2011vesta}.
    }
    \label{Figure2}
\end{figure*}

The polaron formation energy is estimated to be $ E_{\rm{pol}} = -0.4~\mathrm{eV}$, indicating that the localized polaronic state is energetically favorable and thus stable within the MnPS$_3$ monolayer.
Figure~\ref{Figure1}(a) shows the spin-polarized projected density of states (DOS) for the polaronic MnPS$_3$ supercell. The Mn $d$-states are depicted in orange, while the $p$-states of P and S are shown in purple and pink, respectively. The DOS exhibits a localized peak within the energy gap, which can be attributed to the additional polaronic charge. This peak receives contributions from all three types of orbital characters, with the $p$–S orbitals contributing the most significantly. 
The spin-polarized DOS for the delocalized solution is shown in Figure~\ref{fig:A1} displaying the occupation of the lower conduction states by the additional electron.

Figure~\ref{Figure1}(b) displays a comparison between the spin-polarized band structure of the pristine neutral MnPS$_3$ monolayer and the one corresponding to the polaronic MnPS$_3$ monolayer. 
In the neutral case, the AFM configuration exhibits spin degeneracy and an energy gap of approximately 2.5 eV. Upon adding an extra localized charge, a state that originally belonged to the conduction band becomes occupied and is shifted below the Fermi energy, giving rise to a flat band in the spin-up channel, corresponding to the localized peak in Figure~\ref{Figure1}(a). It can be also observed that the presence of the polaron partially lifts the antiferromagnetic spin degeneracy. This effect originates from its unpaired magnetic moment of about $m_s=0.6~ \mu_{\rm{B}}$, delocalized over the three elements, which induces ferromagnetic exchange among the states lying closest in energy to the polaronic state, $i.e.$, the highest occupied and lowest unoccupied states.
Figure~\ref{Figure2} shows the charge density associated with the polaronic state, revealing its orbital character atop the atomic structure.
The polaronic charge is localized within a small region of the material yet extends over several atoms; centered on a P atom, it exhibits trigonal symmetry, with lobes oriented along the P–S bonds and terminating between alternating Mn–Mn distances. Therefore, the presence of the polaron locally breaks the hexagonal symmetry of the magnetic Mn sublattice.

\subsection{Polaron concentration}
\begin{figure*}[t]
    \centering
    \includegraphics[width=\textwidth]{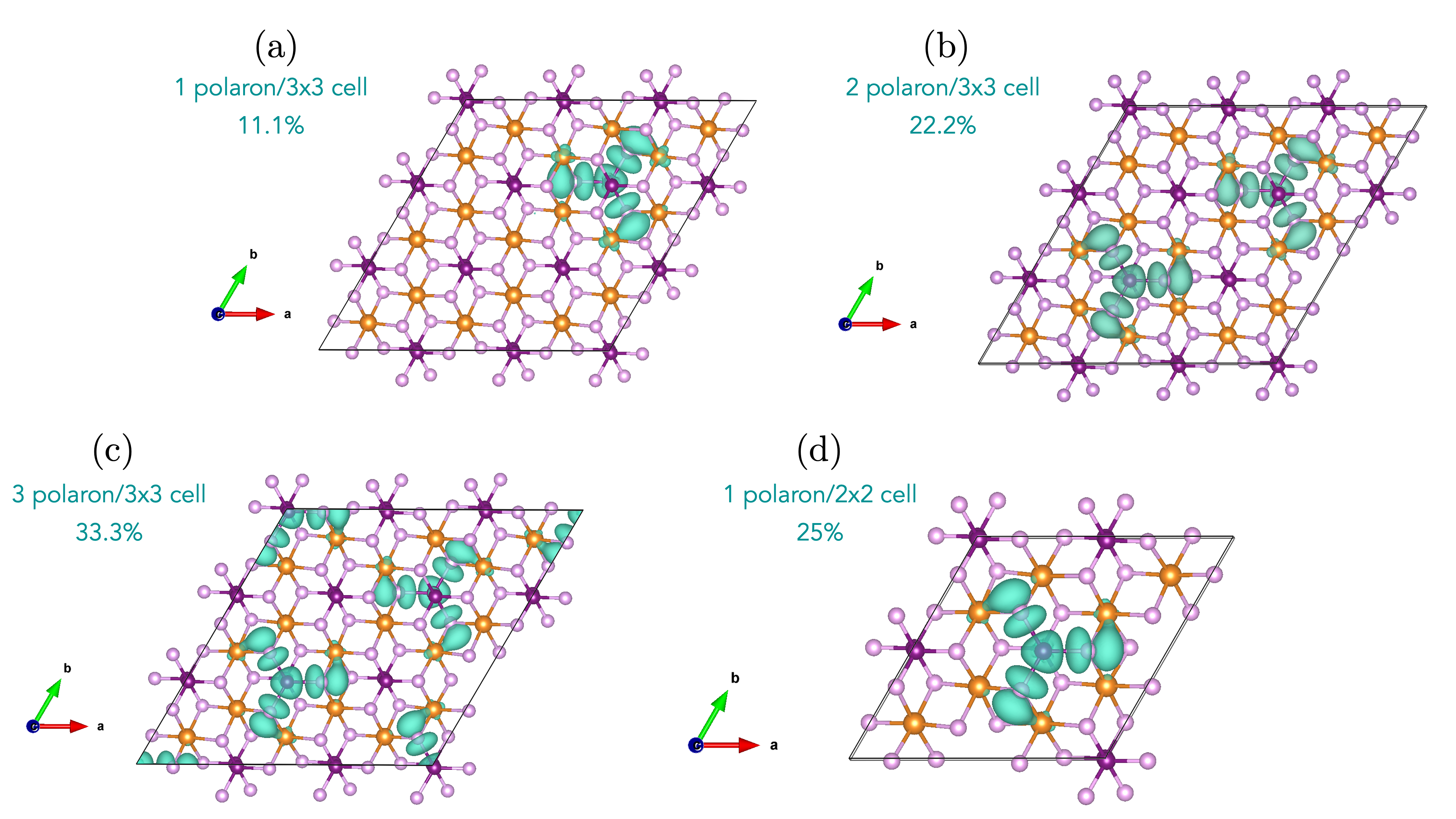}
    \caption{Polarons in MnPS$_3$ monolayer at different concentrations.
    (a) one polaron in a $\sqrt{3}\times\sqrt{3}$ supercell, corresponding to an $11.1\%$ concentration; (b) two polarons in the same $\sqrt{3}\times\sqrt{3}$ supercell, yielding a $22.2\%$ concentration; (c) three polarons in the $\sqrt{3}\times\sqrt{3}$ supercell, corresponding to $33.3\%$, and (d) one polaron in a smaller $2\times2$ supercell, corresponding to a $25\%$ concentration. 
    }
    \label{Figure:concentration}
\end{figure*}

To evaluate the concentration dependence of polaron stability, we examined additional configurations by varying both the supercell size and the number of polarons within the same supercell, spanning concentrations from $11.1\%$ to $50\%$. Our results indicate that polarons remain stable up to a concentration of $33.3\%$. Figure~\ref{Figure:concentration} summarizes the investigated cases: $11.1\%$, $22.2\%$, $25\%$, and $33.3\%$. At higher concentration ($50\%$, corresponding to one polaron in a $1\times2$ supercell, not shown), the localized solution becomes unstable and the excess charge delocalizes. Notably, in all cases where localization occurs, both the orbital occupation and the geometry of the polarons remain consistent with those of the isolated polaron.

\subsection{Exchange tensor}

To assess the effect of the diluted polaron ($i.e.$ 11.1~\% concentration) on the magnetic interactions, we compute the exchange tensor both with and without the the polaronic extra charge. The four-state method was applied by selecting the central pair of magnetic Mn atoms within the supercell, as all Mn sites are symmetrically equivalent. Throughout the calculations, the magnetic background was kept in the AFM configuration, and we used a spin moment $S=5/2$ for the Mn atoms.

\subsubsection{Exchange tensor of bare MnP$S_3$ monolayer}


We first discuss the results for the pristine MnPS$_3$ monolayer without a polaron.
The diagonal elements of the $J_{ij}$ tensor, $J^{xx} = 0.92~\text{meV}$, $J^{yy} = 0.94~\text{meV}$, and $J^{zz} = 0.93~\text{meV}$, differ by less than 3\%, confirming that the exchange interaction is almost perfectly isotropic.
Consequently, the nearest-neighbor (n.n.) exchange interaction is given by $J_1 = \frac{1}{3} \mathrm{Tr}(J_{ij}) = 0.93$~meV. 
Within the standard Heisenberg spin Hamiltonian, the positive value of sign of $J_1$ corresponds to an AFM coupling between n.n. Mn atoms. 
This result is consistent with recent reports, from which we infer 
\( J_1 \sim -0.96~\mathrm{meV} \) based on the reported value of 
$J_1 S^2$ for \( U = 5~\mathrm{eV} \) 
\cite{olsen2021magnetic}, noting that the opposite sign convention for the Heisenberg Hamiltonian is adopted in the literature.\\


\begin{table}[b!]
\centering
\begin{tabular}{p{2.5cm} c c c c c}
\toprule
\parbox[c]{2.5cm}{\centering System} & Mn pair & Distance & $J^{xx}$ & $J^{yy}$ & $J^{zz}$ \\
\midrule
\parbox[c]{2.5cm}{\centering Non-polaronic \\ (undistorted)} 
 & Mn--Mn & 3.548 & 0.92  & 0.94  & 0.93 \\
\midrule
\multirow{2}{*}{\parbox[c]{2.5cm}{\centering Polaronic \\ (distorted)}} 
 & Mn$_1$--Mn$_2$ & 3.587 & 0.61 &-0.07  & -0.07 \\
 & Mn$_2$--Mn$_3$ & 3.475 & 1.07 & 0.44  &0.44  \\
\midrule
\multirow{2}{*}{\parbox[c]{2.5cm}{\centering Non-polaronic \\ (distorted)}} 
 & Mn$_1$--Mn$_2$ & 3.587 & 0.84 & 0.86  & 0.86  \\
 & Mn$_2$--Mn$_3$ & 3.475 & 1.27 & 1.29 & 1.30 \\
\bottomrule
\end{tabular}
\caption{Mn--Mn distances and exchange parameters for non-polaronic, polaronic, and distorted MnPS$_3$. Distances in \AA, exchange interactions in meV.}
\label{Table1}
\end{table}


\subsubsection{Exchange tensor of polaronic MnP$S_3$ monolayer} 

The introduction of a polaron into the material causes a local distortion of the lattice, altering the distance between Mn atoms and generating symmetry-inequivalent pairs. The displacement directions of the Mn atoms is indicated in Figure~\ref{Figure2} by blue arrows, whose lengths are proportional to the corresponding displacement magnitudes, $\approx$ 0.05~\AA.
The computed interatomic distances for the investigated cases are given in Table~\ref{Table1}.

To assess the effect of the polaron on the exchange tensor we repeated the non-collinear calculations by selecting two Mn pairs, Mn$_1$–Mn$_2$ and Mn$_2$–Mn$_3$, which become symmetry-inequivalent due to the presence of the trigonal-symmetric localized polaronic charge, as illustrated in Figure~\ref{Figure2}. The resulting exchange tensors remain diagonal, with the first pair exhibiting components $J_{1,2}^{xx} = 0.61$~meV and $J_{1,2}^{yy} = J_{1,2}^{zz} = -0.07$~meV, while the second pair shows $J_{2,3}^{xx} = 1.07$~meV and $J_{2,3}^{yy} = J_{2,3}^{zz} = 0.44$~meV. 
These results indicate that the presence of the polaron not only modifies the magnitude of the exchange tensor components but also introduces anisotropy in the exchange interaction, as evidenced by the disparity between the $xx$ component and the $yy = zz$ components. Moreover, the $J_{1,2}$ exchange further exhibits a polaron-induced sign change in the $yy$ and $zz$ components, indicating a weak ferromagnetic interaction. Nevertheless, these components remain significantly smaller than the dominant $xx$ term; consequently, for both Mn pairs, the overall exchange interaction is governed primarily by an AFM coupling. A closer inspection of Table~B2 reveals that total energy calculations within the four-states method identify the $\downarrow\downarrow$ configuration for the $J_{xx}$ component as the ground state for both Mn pairs. In contrast, without the additional localized electron (pristine and distorted cases), the antiparallel spin configurations are always energetically preferred. This behavior suggests that the polaron induces a spin flip, thereby modifying the local magnetic configuration. Thus, although the pure magnetic exchange intrinsically favors AFM coupling, additional polaron-related effects act to destabilize this configuration and promote spin flipping. A more detailed analysis of the microscopic origin of these competing interactions will be the subject of future investigations.

To better understand the origin of the asymmetric exchange, we decoupled the two primary effects introduced by the polaron: the excess electronic charge and the associated crystal field distortions. This was achieved by performing additional non-collinear calculations of the $J_{ij}$ tensor on the polaron-distorted structure, but in the absence of the extra electron.
The computed values are 
\( J_{1,2}^{xx} = 0.84~\mathrm{meV} \), 
\( J_{1,2}^{yy} = 0.86~\mathrm{meV} \), and 
\( J_{1,2}^{zz} = 0.86~\mathrm{meV} \), as well as 
\( J_{2,3}^{xx} = 1.27~\mathrm{meV} \), 
\( J_{2,3}^{yy} = 1.29~\mathrm{meV} \), and 
\( J_{2,3}^{zz} = 1.30~\mathrm{meV} \). 
Since these components differ by less than 3\% of the average value, 
this confirms a restored isotropic exchange interactions, and demonstrates that the magnetic anisotropy observed in the polaronic system vanishes in the absence of the excess charge.
This restored isotropy indicates that the anisotropic exchange arises primarily from the presence of the localized polaronic charge, rather than from lattice distortions alone. Nevertheless, the distortions still affect the magnitude of the exchange interactions: $J_1$ for the Mn$_1$–Mn$_2$ pair is slightly reduced compared to the pristine MnPS$_3$ monolayer, while $J_1$ for the Mn$_2$–Mn$_3$ pair is enhanced—correlating with the previously discussed changes in Mn–Mn bond lengths. Table~\ref{Table1} summarizes the exchange interactions and bond lengths for the Mn pairs in the three configurations considered: pristine MnPS$_3$, polaronic (including the extra charge and lattice distortions), and lattice-distorted without the excess charge.\\

\section{Conclusions}

In conclusion, we have shown that a single electron polaron can energetically stabilize in a monolayer of the semiconducting antiferromagnet MnPS$_3$, resulting in a localized in-gap state. The polaron exhibits a complex orbital character, centered on a P atom with a trigonal symmetry that extends over neighboring S and Mn atoms. To assess its influence on magnetic properties, we computed the magnetic exchange interactions in full tensorial form $J_{ij}$. Our results reveal that the presence of the polaron locally breaks the symmetry of the pristine antiferromagnetic configuration, introducing anisotropy into the exchange coupling. Importantly, this anisotropic behavior is primarily driven by the excess localized electronic charge, whereas the associated lattice distortions mainly modulate the magnitude of the isotropic exchange terms.
These findings highlight a promising strategy for local control of magnetism in 2D systems: by exploiting polarons as agents for magnetic modification. Furthermore, the polaron-induced anisotropy in the exchange interactions is expected to influence the magnon spectrum, potentially modifying spin-wave energies and dispersions in ways that could be probed in neutron scattering experiments, thereby demonstrating a direct link between localized charge, lattice distortions, and collective spin excitations. This underlines the importance for further exploration of polaronic 2D magnets as potential platforms for spintronic device applications.\\ 
 
\textit{Acknowledgments---}
We are grateful to Gustav Bihlmayer and Dario Fiore Mosca for their valuable insights. J.P.C. and J.B. also thank Eduardo Mendive Tapia for the engaging discussions. The authors acknowledge the Erwin Schrödinger Institute (ESI) for hosting the ESI-PsiK workshop "Spin–orbit Entangled Quantum Magnetism". \\
This research was supported by the Austrian Science Fund (FWF) "Spin-orbit entangled anharmonic polarons" 10.55776/PIN5456724.
The computational results presented have been achieved using the Austrian Scientific Computing (ASC).\\

\textit{Data availability---}
The data supporting the
findings are available within the manuscript.

\bibliographystyle{apsrev4-2}
\bibliography{bib}
\appendix
\setcounter{figure}{0}
\setcounter{table}{0}
\renewcommand{\thefigure}{A\arabic{figure}}
\renewcommand{\thetable}{B\arabic{table}}
\section{Density of States with a delocalized extra charge}
The total DOS of the delocalized solution is calculated in the MnPS$_3$ $3\times3\times1$ supercell with an extra electron, which occupies the lowest conduction band states.
 \begin{figure}[h!]
    \centering
    \includegraphics[scale=0.4]{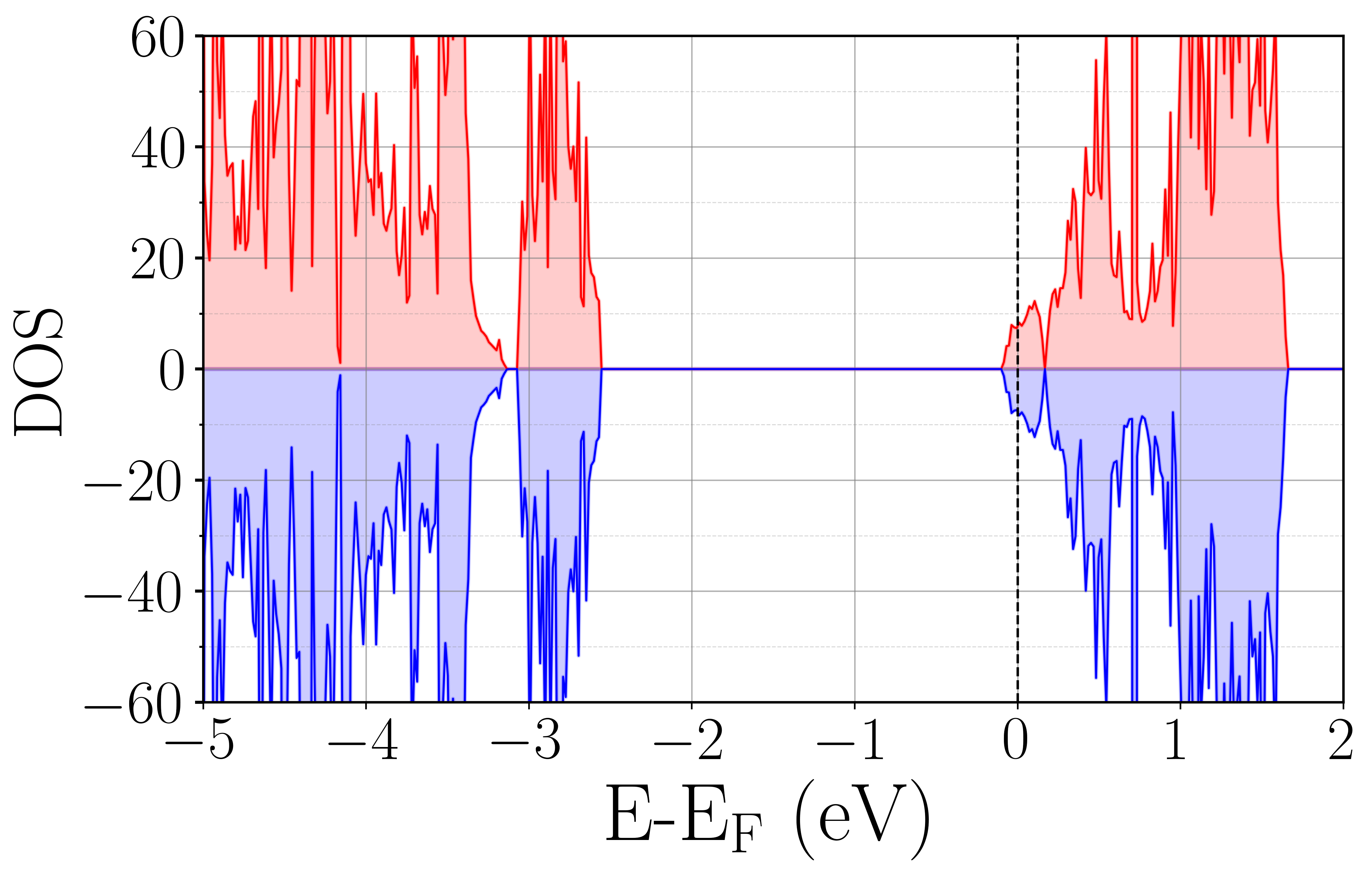}
    \caption{Spin-polarized DOS of the doped $3\times3\times1$ MnPS$_3$ monolayer in the delocalized charge configuration. The upper panel displays the majority spin channel (red), while the lower panel corresponds to the minority spin channel (blue).}
    \label{fig:A1}
\end{figure}

\newpage
\onecolumngrid
\section{Total energy values for the four-state method}
\vspace{-0.7cm}
\begin{table}[h]
\caption{Total energies (eV) of pristine MnPS$_3$ for the Mn$_1$–Mn$_2$ pair, for each $J^{ab}_{ij}$ component.}
\label{tab:fourstate}
\begin{ruledtabular}
\begin{tabular}{c cccc}
$ab$ & $\uparrow\uparrow$ & $\downarrow\downarrow$ & $\uparrow\downarrow$ & $\downarrow\uparrow$ \\
\hline
$xx$ & $-493.2407$ & $-493.2407$ & $-493.2842$ & $-493.2202$ \\
$xy$ & $-493.2600$ & $-493.2279$ & $-493.2600$ & $-493.2279$ \\
$xz$ & $-493.2600$ & $-493.2279$ & $-493.2599$ & $-493.2279$ \\
$yx$ & $-493.2279$ & $-493.2600$ & $-493.2600$ & $-493.2279$ \\
$yy$ & $-493.2371$ & $-493.2371$ & $-493.2488$ & $-493.2488$ \\
$yz$ & $-493.2425$ & $-493.2425$ & $-493.2425$ & $-493.2422$ \\
$zx$ & $-493.2280$ & $-493.2601$ & $-493.2602$ & $-493.2279$ \\
$zy$ & $-493.2426$ & $-493.2425$ & $-493.2425$ & $-493.2425$ \\
$zz$ & $-493.2374$ & $-493.2374$ & $-493.2490$ & $-493.2491$ \\
\end{tabular}
\end{ruledtabular}
\end{table}
\vspace{-0.7cm}
\hfill
\begin{table}[h]
\caption{Total energies (eV) of polaronic MnPS$_3$ for the Mn$_1$–Mn$_2$ and Mn$_2$–Mn$_3$ pairs, for each $J^{ab}_{ij}$ component.}
\label{tab:polaronic}
\begin{ruledtabular}
\begin{tabular}{c cccc cccc}
 & \multicolumn{4}{c}{Mn$_1$–Mn$_2$} & \multicolumn{4}{c}{Mn$_2$–Mn$_3$} \\
$ab $ & $\uparrow\uparrow$ & $\downarrow\downarrow$ & $\uparrow\downarrow$ & $\downarrow\uparrow$ &
$\uparrow\uparrow$ & $\downarrow\downarrow$ & $\uparrow\downarrow$ & $\downarrow\uparrow$ \\
\hline
$xx$ & $-494.1799$ & $-494.2971$ & $-494.2773$ & $-494.2149$ & $-494.1787$ & $-494.2971$ & $-494.2773$ & $-494.2253$ \\
$xy$ & $-494.2258$ & $-494.2531$ & $-494.2258$ & $-494.2531$ & $-494.2253$ & $-494.2577$ & $-494.2253$ & $-494.2577$ \\
$xz$ & $-494.2258$ & $-494.2531$ & $-494.2258$ & $-494.2531$ & $-494.2253$ & $-494.2577$ & $-494.2253$ & $-494.2577$ \\
$yx$ & $-494.1933$ & $-494.2829$ & $-494.2829$ & $-494.1933$ & $-494.1985$ & $-494.2829$ & $-494.2829$ & $-494.1985$ \\
$yy$ & $-494.2377$ & $-494.2377$ & $-494.2369$ & $-494.2369$ & $-494.2372$ & $-494.2372$ & $-494.2426$ & $-494.2426$ \\
$yz$ & $-494.2364$ & $-494.2365$ & $-494.2365$ & $-494.2365$ & $-494.2390$ & $-494.2389$ & $-494.2390$ & $-494.2390$ \\
$zx$ & $-494.1933$ & $-494.2829$ & $-494.2829$ & $-494.1933$ & $-494.1985$ & $-494.2829$ & $-494.2829$ & $-494.1985$ \\
$zy$ & $-494.2364$ & $-494.2364$ & $-494.2365$ & $-494.2364$ & $-494.2390$ & $-494.2390$ & $-494.2390$ & $-494.2390$ \\
$zz$ & $-494.2378$ & $-494.2378$ & $-494.2369$ & $-494.2369$ & $-494.2372$ & $-494.2372$ & $-494.2426$ & $-494.2426$ \\
\end{tabular}
\end{ruledtabular}
\end{table}
\vspace{-0.7cm}
\begin{table}[h]
\caption{Total energies (eV) in distorted MnPS$_3$ for the Mn$_1$–Mn$_2$ and Mn$_2$–Mn$_3$ pairs, for each $J^{ab}_{ij}$ component.}
\label{tab:distorted}
\begin{ruledtabular}
\begin{tabular}{c cccc cccc}
 & \multicolumn{4}{c}{Mn$_1$–Mn$_2$} & \multicolumn{4}{c}{Mn$_2$–Mn$_3$} \\
$ab$  & $\uparrow\uparrow$ & $\downarrow\downarrow$ & $\uparrow\downarrow$ & $\downarrow\uparrow$ &
$\uparrow\uparrow$ & $\downarrow\downarrow$ & $\uparrow\downarrow$ & $\downarrow\uparrow$ \\
\hline
$xx$ & $-492.3477$ & $-492.3456$ & $-492.3932$ & $-492.3210$ & $-492.3470$ & $-492.3456$ & $-492.3932$ & $-492.3311$ \\
$xy$ & $-492.3678$ & $-492.3306$ & $-492.3678$ & $-492.3306$ & $-492.3674$ & $-492.3357$ & $-492.3674$ & $-492.3357$ \\
$xz$ & $-492.3677$ & $-492.3306$ & $-492.3678$ & $-492.3306$ & $-492.3674$ & $-492.3357$ & $-492.3674$ & $-492.3357$ \\
$yx$ & $-492.3315$ & $-492.3667$ & $-492.3667$ & $-492.3315$ & $-492.3363$ & $-492.3668$ & $-492.3668$ & $-492.3363$ \\
$yy$ & $-492.3425$ & $-492.3425$ & $-492.3532$ & $-492.3532$ & $-492.3428$ & $-492.3428$ & $-492.3590$ & $-492.3590$ \\
$yz$ & $-492.3472$ & $-492.3472$ & $-492.3472$ & $-492.3472$ & $-492.3499$ & $-492.3499$ & $-492.3499$ & $-492.3499$ \\
$zx$ & $-492.3316$ & $-492.3667$ & $-492.3668$ & $-492.3315$ & $-492.3363$ & $-492.3667$ & $-492.3667$ & $-492.3363$ \\
$zy$ & $-492.3472$ & $-492.3472$ & $-492.3472$ & $-492.3472$ & $-492.3499$ & $-492.3499$ & $-492.3499$ & $-492.3499$ \\
$zz$ & $-492.3425$ & $-492.3425$ & $-492.3532$ & $-492.3532$ & $-492.3428$ & $-492.3428$ & $-492.3590$ & $-492.3590$ \\
\end{tabular}
\end{ruledtabular}
\end{table}

\end{document}